\documentclass[12pt]{article}
\usepackage{amsmath}
\usepackage{amssymb}
\usepackage{citesort}
\usepackage{graphicx}
\usepackage[nomarkers]{endfloat}
\usepackage{pstricks}
\begin{document}
\setcounter{page}{0}
\thispagestyle{empty}
\begin{flushright}
{\small BARI-TH 395/00}
\end{flushright}
\vspace*{2.5cm}
\begin{center}
{\large\bf Comment on \\
 "No Primordial Magnetic Field from Domain Walls"}
\end{center}

\vspace*{2cm}

\renewcommand{\thefootnote}{\fnsymbol{footnote}}

\begin{center}
{ P. Cea$^{1,2,}$\protect\footnote{Electronic address: {\tt
Cea@bari.infn.it}},
and
L. Tedesco$^{1,2,}$\protect\footnote{Electronic address: {\tt
Tedesco@bari.infn.it}} \\[0.5cm] $^1${\em Dipartimento di Fisica,
Universit\`a di Bari, I-70126 Bari, Italy}\\[0.3cm] $^2${\em INFN
- Sezione di Bari, I-70126 Bari, Italy} }
\end{center}

\vspace*{0.5cm}

\begin{center}
{
October, 2000 }
\end{center}

\vspace*{1.0cm}

\renewcommand{\abstractname}{\normalsize Abstract}
\begin{abstract}
We comment on the recent preprint hep-ph/0007123 by
M.B.~Voloshin, claiming that domain walls are diamagnetic.
We show that the results presented therein  are based on an
incorrect treatment of the zero mode contribution
to the vacuum energy density.  We also shown that the correct treatment
leads to the conclusion that domain walls are ferromagnetic, and
can generate a primordial magnetic field.
\end{abstract}
\vfill
\newpage

In Ref.~\cite{Cea:1999} we suggested that the spontaneous
generation of uniform magnetic condensate in
$QED_3$~\cite{Cea:1985},~\cite{Hosotani:93},~\cite{Cea:2000} could
give rise to ferromagnetic domain walls at the electroweak phase
transition. Moreover, we suggested that these domain walls
generate a magnetic field at the electroweak scale which can be
relevant for the generation of the {\it primordial}
magnetic field. \\
In a recent paper~\cite{Voloshin:2000}, Voloshin claims that the
generation of primordial magnetic fields by domain walls is very
unlikely. In particular, the author of Ref.~\cite{Voloshin:2000}
points out that massive $(2+1)$ dimensional fermions bound to a
domain wall behave diamagnetically rather than ferromagnetically.
In addition, Voloshin affirms
that our previous claim of ferromagnetism is based on a incorrect
formula in Ref.~\cite{Cea:1985}  for the magnetic field
contribution to the fermionic vacuum free energy. \\
The aim of the present comment is twofold. First, we show  that
the results of ~\cite{Voloshin:2000} derive from an incorrect
treatment of the fermionic zero mode contribution to the zero
temperature vacuum energy density. Secondly, we show that the
finite temperature calculation of Ref.~\cite{Voloshin:2000} agrees
with our previous findings~\cite{Cea:2000},~\cite{Cea:1998}.
Moreover, if one takes into account the correct definition of an
Abelian magnetic field in presence of a varying scalar field
condensate, then it turns out that the classical magnetic energy
is proportional to the area of the wall. This last result supports
our previous proposal in Ref.~\cite{Cea:1999}. \\
Let us consider the zero temperature vacuum energy density due to
massive planar fermions in presence of a constant magnetic field:
\begin{equation}
\label{Eq.1}
E_{\rm vac}= - \frac{e \, B}{2 \, \pi} \,\sum_{n=0}^\infty
\sqrt{2 \, e \, B \, n +m^2}~.
\end{equation}
The infinite sum in Eq.~(\ref{Eq.1}) needs to be regularized.
Among the possible gauge-invariant choices, we employed the
Schwinger proper-time regularization scheme. However, any gauge
invariant regularization gives physically sensible results.
Following Ref.~\cite{Voloshin:2000} we regularize the sum by means
of the gauge invariant cut-off $\exp(-\epsilon \, E_n^2)$, where
$\epsilon$ is the regulator parameter and $E_n$ is the energy of
the levels. We get:
\begin{equation}
\label{Eq.2}
E_{\rm vac}^{(r)}(B)= -  \frac {e \, B}  {2 \pi}
\,\sum_{n=0}^\infty \sqrt{2 \, e \, B \, n +m^2} \, \exp(-\epsilon
\, 2 \, e \, B \, n - \epsilon \, m^2)~.
\end{equation}
To evaluate the sum in Eq.~(\ref{Eq.2}), Voloshin uses the
Poisson's summation formula:
\begin{equation}
\label{Eq.3}
\sum_{n=0}^\infty \, f(n) = \int_{\delta}^\infty \, f(x) \,
\sum_{n=-\infty}^\infty \, \delta(x-n) \, dx =
\sum_{k=-\infty}^\infty \, \int_{\delta}^\infty \, f(x) \, \exp( 2
\, \pi \, i \, k \, x) \, dx~~,
\end{equation}
where $\delta$ is such that $-1 < \delta <0$. Now, observing that
the summand in Eq.~(\ref{Eq.2}) is non singular at $n=0$, Voloshin
assumes that one can set $\delta=0$. However it is easy to see
that this last assumption is not valid. Indeed, if we consider the
$n=0$ term in Eq.~(\ref{Eq.3}) :
\begin{equation}
\label{Eq.4}
I_\delta \equiv \int_{\delta}^\infty \, f(x) \, \delta(x) \, dx \;
,
\end{equation}
we see that $\lim_{\delta \rightarrow 0^-} I_\delta = f(0)$, while
$\lim_{\delta \rightarrow 0^+} I_\delta = 0$. It is now evident
that the limit $\delta \rightarrow 0$ is not harmless. Indeed, the
Voloshin's procedure is equivalent to ignore the zero
mode contribution. \\
The correct procedure can be obtained as follows. First one must
isolate the zero mode contribution and, then, one can apply the
Poisson's summation formula to the $n \geq 1$ modes. In this way,
in the weak magnetic field region we get for the vacuum energy
density:
\begin{eqnarray}
\label{Eq.5}
 E_{\rm vac}(B)-E_{\rm vac}(0)
& = &
 -  \frac {eB} {4 \pi} |m| + \frac {e^2 \, B^2}
 { 2 \, \pi^3 \, |m|} \sum_{p=0}^\infty \frac {(-1)^p \, \zeta(2 \, p +2)
\, \Gamma(\frac {3} {2}) } { \Gamma (\frac {1}  {2} - 2 \, p)} \,
\left ( \frac {e \, B} { \pi \, m^2} \right )^{2p}
 \nonumber \\
 &  &
  = - \frac {eB} {4 \pi} |m| + \frac {e^2 \, B^2} { 24 \, \pi \, |m|} +
\ldots ~.
\end{eqnarray}
This last equation agrees with the results obtained in
Ref.~\cite{Cea:1985}, and differs from Eq.~(14) of
Ref.~\cite{Voloshin:2000} in the negative term, linear in the magnetic
field. Obviously, it is the linear term which gives rise to
the spontaneous magnetic condensation in $QED_3$.
We would like to note that, in the case of the thermal
corrections, Voloshin correctly separates the zero mode
contribution in the free energy calculation; therefore, it is
surprising that this is not done also at zero temperature in
~\cite{Voloshin:2000}. \\
Equation~(\ref{Eq.5}) shows that the  results reported in
Ref.~\cite{Voloshin:2000}, corrected to include the zero mode
contributions,  fully support the spontaneous generation of a
uniform magnetic condensate in $QED_3$ with massive fermions.
Moreover, it is worthwhile to stress that even the free energy
higher temperature study of Ref.~\cite{Voloshin:2000} is in
agreement with the one reported in
Ref.~\cite{Cea:2000},~\cite{Cea:1998}. Indeed, if one takes into
account the missing linear term in the Voloshin's zero temperature
energy density, it easy to see that Eq.(18) of
Ref.~\cite{Voloshin:2000} is in accordance with our previous
conclusion that the thermal corrections, even at infinite
temperature, do not modify the spontaneous generation of the
magnetic condensate. \\
Let us now comment on the Voloshin's claim that the domain walls
cannot be a source of the primordial magnetic field. This
statement in ~\cite{Voloshin:2000} is based  on the argument that
the classical energy of the induced magnetic field is proportional
to the volume of the box, while the contribution due to the
fermion modes localized on the wall scales with the area of the
wall. In this way Voloshin's total energy of the system is given
by:
\begin{equation}
\label{Eq.6}
 E(B)=L^3 \, \frac {B^2 } {2} + L^2 \, f(B) \;,
\end{equation}
where $L$ is the linear size of the system. It turns out that the
above equation forgets completely the non Abelian nature of the
induced magnetic field. Indeed, in the case of varying scalar
field condensate the correct definition of Abelian electromagnetic
field is given by the t'Hooft's Abelian
projection~\cite{'tHooft:1974}. Taking into account that in our
model the Abelian part of the Abelian projected magnetic field is
induced by the fermionic modes localized on the wall, it is easy
to see that the Abelian projected magnetic field vanishes in the
regions where the scalar condensate is constant. So that, in
general the appropriate expression for the Abelian magnetic field
can be different from zero only in the regions where the scalar
condensate varies. In our model the magnetic field is localized in
a region of the order of the wall thickness $\Delta$. Thus
Eq.~(\ref{Eq.6}) is replaced by:
\begin{equation}
\label{Eq.7}
 E(B)=L^2 \, \Delta \, \frac {B^2}  { 2} + L^2 \, f(B) \;.
\end{equation}
which led to the conclusions of Ref.~\cite{Cea:1999}. \\
In conclusion, it is worthwhile to stress that in the realistic
case where the domain wall interacts with the plasma, the magnetic
field penetrates into the plasma over a distance of the order of
the penetration length $\lambda$ which is about an order of
magnitude greater than $\Delta $. It follows that the estimate in
Ref.~\cite{Cea:1999} of the induced magnetic field at the
electroweak scale $ B^* \, \simeq \, 5 \, 10^{24} \, Gauss$ is
reduced by a factor $\sqrt{\frac {\Delta} {\lambda}} \, \sim \,
0.3$  which, however, is still of cosmological interest.
%
%
\vspace*{4.5cm}
\end{document}